\documentclass[11pt, a4paper]{article}
\usepackage{amsmath}

\newcommand{\benumerate}{\begin{enumerate}}
\newcommand{\eenumerate}{\end{enumerate}}

\newcommand{\bitemize}{\begin{itemize}}

\newcommand{\eitemize}{\end{itemize}}
\newcommand{\ep}{\epsilon}
\newcommand{\vphi}{\varphi}

\begin{document}

\title{Dispersive deformations of hydrodynamic reductions of 2D
dispersionless integrable systems}
\author{E.V. Ferapontov, A. Moro}
    \date{}
    \maketitle
    \vspace{-7mm}
\begin{center}
Department of Mathematical Sciences \\ Loughborough University \\
Loughborough, Leicestershire LE11 3TU \\ United Kingdom \\[2ex]
e-mails: \\[1ex] \texttt{E.V.Ferapontov@lboro.ac.uk}\\
\texttt{A.Moro@lboro.ac.uk}
\end{center}

\bigskip

\begin{abstract}

We demonstrate that hydrodynamic reductions of  dispersionless
integrable systems in $2+1$ dimensions, such as the dispersionless
Kadomtsev-Petviashvili (dKP) and dispersionless Toda lattice (dTl)
equations, can be deformed into reductions of the corresponding
dispersive counterparts. Modulo the Miura group, such deformations
are unique. The requirement that {\it any} hydrodynamic reduction
possesses a deformation of this kind   imposes strong constraints
on the structure of dispersive terms,  suggesting an alternative
approach to the integrability in $2+1$  dimensions.

\bigskip


\bigskip

\end{abstract}

\newpage

\section{Introduction}

The Kadomtsev-Petviashvili (KP) equation,
\begin{equation}
\label{kp} (u_{t} -  u u_{x} - \frac{\epsilon^{2}}{12} u_{xxx} )_x=u_{yy},
\end{equation}
arises in mathematical physics as a  two-dimensional generalization of the KdV equation. Although its integrability aspects have been thoroughly investigated in the literature, we believe that one important property has been overlooked, namely, that  Eq. (\ref{kp})  can be decoupled into a pair of consistent $(1+1)$-dimensional equations in a {\it continuum} of ways. These decouplings can be obtained as  deformations of hydrodynamic reductions of its dispersionless limit,  known as the dKP equation,
\begin{equation}
\label{dkp} (u_{t} -  u u_{x} )_x=u_{yy}.
\end{equation}
We point out that  Eq. (\ref{dkp}), also known as the Khokhlov-Zabolotskaya equation \cite{KZ}, is of interest in its own, arising in non-linear acoustics, gas dynamics and differential geometry.  A key property of  the dKP equation is the existence of  $n$-phase solutions of the form
\begin{equation}
u=u(R^1, ..., R^n),
\label{u}
\end{equation}
where the `phases'  $R^i(x, y, t)$ are governed by a pair of commuting hydrodynamic type systems
\begin{equation}
 R^i_y=\mu^i(R) R^i_x,  ~~~~ R^i_t=\lambda^i(R) R^i_x.
\label{Ri}
\end{equation}
Here  $\lambda^i=(\mu^i)^2+u$, while $\mu^i$ and $u$ satisfy the so-called Gibbons-Tsarev equations,
\begin{equation}
\partial_j\mu^i=\frac{\partial_j u}{\mu^j-\mu^i}, ~~~
\partial_i\partial_ju=2\frac{\partial_iu\partial_ju}{(\mu^j-\mu^i)^2},
\label{GT}
\end{equation}
$i\ne j$, $\partial_i=\partial/\partial R^i$, which were first derived in \cite{GibTsa96, GibTsa99} in the
context of hydrodynamic reductions of Benney's moment equations. Thus, Eqs. (\ref{Ri}) can be viewed as a decomposition of the $(2+1)$-dimensional Eq. (\ref{dkp}) into a pair of $(1+1)$-dimensional hydrodynamic type systems. We will refer to Eqs. (\ref{Ri}) as  hydrodynamic reductions of dKP. These reductions have been extensively studied in the literature, see e.g. \cite{Gibb94, Fer4, Ma} and references therein. In particular, in the one-component case Eqs. (\ref{GT}) become vacuous, and without any loss of generality one can set $u(R)=R$ where $R(x, y, t)$ satisfies a pair of Hopf-type equations
\begin{gather}
\label{R}
\begin{aligned}
R_{y} = \mu R_{x} , ~~~~ R_{t} =(\mu^{2} + R) R_{x};
\end{aligned}
\end{gather}
here $\mu(R)$ is an arbitrary function. We recall that the general solution of Eqs. (\ref{R}) is given by the implicit formula $f(R)=x+\mu y+(\mu^2+R)t$, which implies that the level surfaces $R$=const are planes. Solutions of this type are known as planar simple waves.

Our main observation is that {\it all} hydrodynamic reductions (\ref{Ri})  can be deformed into  reductions of the full KP equation by adding appropriate dispersive terms which are {\it polynomial} in the $x$-derivatives of $R^i$. Up to Miura-type transformations, such deformations are unique. Moreover, the calculation of dispersive corrections  is an entirely algebraic procedure which does not require  solving differential equations. In the one-component case one obtains the following deformation of Eqs. (\ref{R}):
\begin{gather}
\label{R_Def}
\begin{aligned}
R_{y} =& \mu R_{x} \\
& +\frac{\epsilon^{2}}{12} \left(\mu' R_{xx} +\frac{1}{2}
(\mu''- (\mu')^3) R_{x}^2 \right)_x + O(\epsilon^{4}),\\
R_{t} =& (\mu^{2} + R) R_{x} \\
&+\frac{\epsilon^{2}}{12} \left( (2\mu \mu'+1)R_{xx}+( \mu \mu''-\mu (\mu')^3+(\mu')^2/2)
R_{x}^2 \right)_x  + O(\epsilon^{4}),
\end{aligned}
\end{gather}
see  Sect. 2.1 for more details. Notice that the relation $u=R$, which now solves the full KP equation,  remains undeformed: this can always be assumed modulo the Miura group. Eqs. (\ref{R_Def}) can be viewed as a decomposition of KP  into a pair of commuting $(1+1)$-dimensional equations parametrized by an arbitrary function of one variable.  In general, these equations constitute infinite series in  $\epsilon^2$ which can terminate only in exceptional cases (in the one-component situation this happens only when $\mu$=const, in which case KP reduces to KdV). One can show that deformations (\ref{R_Def}) are nontrivial, that is, not reducible to Eqs. (\ref{R}) by a Miura-type transformation.   Formal expansions of the type (\ref{R_Def}) have appeared in \cite{Baikov}, and were thoroughly investigated in a series of publications  \cite{Dub1, Dub2, Dub3, Zhang, Strachan2} in the context of 2D topological field theory. We would like to formulate the following  conjecture:

\medskip

\noindent {\it For any integrable soliton system in $2+1$ dimensions, all hydrodynamic reductions of its dipersionless limit can be deformed into reductions of the original  system.}

\medskip

\noindent Thus, any $(2+1)$-dimensional integrable soliton equation can be decoupled into a pair of compatible $(1+1)$-dimensional equations in an infinity of ways. Deformations of two-component reductions of dKP are discussed in Sect. 2.2 and 2.3.  Another example supporting our conjecture is provided by the Toda lattice system,
\begin{gather}
\label{TL}
\begin{aligned}
 \epsilon u_{y} &= u \; (w(x) - w(x-\epsilon)), \\
\epsilon w_{t} &= u(x+ \epsilon) - u(x),
\end{aligned}
\end{gather}
whose dispersionless limit (dTl equations) assumes  the form
$$
u_{y} = u w_{x}, \qquad w_{t} = u_{x}.
$$
In Sect. 3 we provide explicit formulae for dispersive deformations of one-component reductions of dTl.

The requirement that {\it all} hydrodynamic reductions of a $(2+1)$-dimensional dispersionless integrable system can be deformed into reductions of the corresponding dispersive equation imposes strong constraints on the structure of dispersive terms. As an illustration, let us consider the generalized KP  equation of the form
\begin{gather}
\label{kp_Ext}
\begin{aligned}
\left( u_{t} - u u_{x} + \epsilon (A_{1} u_{xx} + A_{2} u_{x}^{2})
 + \epsilon^{2} (B_{1} u_{xxx} + B_{2} u_{x} u_{xx} + B_{3} u_{x}^{3})\right)_x=
u_{yy},
\end{aligned}
\end{gather}
where $A_{i}$ and $B_{i}$ are certain functions of $u$. Notice that this equation has the same dispersionless limit as the KP equation (\ref{kp}). As demonstrated in Sect. 4.1, the requirement that {\it all} one-component reductions of dKP can be deformed into reductions of (\ref{kp_Ext}) readily implies $A_1=A_2=B_2=B_3=0$ and $ B_1$=const, moreover, to establish this one only needs to perform calculations up to the order $\epsilon^4$. Thus, our procedure uniquely reconstructs the KP equation. Further examples of this type include  BKP/CKP and the `universal' equation (Sect. 4.2 and 4.3, respectively). This naturally leads to the program of classification of $(2+1)$-dimensional integrable soliton equations which can be summarized as follows:

\medskip

\noindent (a) Classify $(2+1)$-dimensional dispersionless integrable systems within various particularly interesting classes.
A number of results in this direction are already available, see e.g. \cite{Bla, Fer4, Fer5, Fer6, Fer7, Fer8, Moro}, etc.
We recall that the integrability of a $(2+1)$-dimensional dispersionless  system is understood as the existence, for any $n$, of an infinity of $n$-component hydrodynamic reductions parametrized by $n$ arbitrary functions of one variable \cite{Fer4}.

\medskip

\noindent (b) Reconstruct possible dispersive terms from the requirement that  {\it all} hydrodynamic reductions of the dispersionless system can be deformed into reductions of the corresponding dispersive equation. We conjecture that any $(2+1)$-dimensional dispersionless integrable system can be deformed in this way (possibly, non-uniquely).

\medskip

This scheme  can be viewed as an alternative to the classical approach to the integrability in $2+1$ dimensions, which starts with a linear dispersive part, say, $u_{tx} - \frac{\epsilon^{2}}{12} u_{xxxx} =u_{yy}$, and reconstructs the allowed  nonlinearity
\cite{Shulman, Mikhailov}.



\section{Deformations of  the dKP reductions}

We will consider separately deformations of one- and two-component reductions of dKP.

\subsection{One-component reductions}

Let us first rewrite both KP and  dKP equations as two-component  systems,
\begin{equation}
\label{KP} u_{t} -  u u_{x} - \frac{\epsilon^{2}}{12} u_{xxx}= w_{y}, \qquad u_{y} = w_{x}
\end{equation}
and
\begin{equation}
\label{dKP} u_{t} -  u u_{x} = w_{y}, \qquad u_{y} = w_{x},
\end{equation}
respectively; such representation simplifies the calculations. One-component (one-phase) reductions of dKP are given by the formula
\begin{equation}
u = R,  \qquad w =w(R),
\label{uw}
\end{equation}
where $w'=\mu$ and $R(x, y, t)$ solves Eqs. (\ref{R}). Dispersive deformations of  Eqs.
(\ref{uw}), (\ref{R}) are sought in the form
\begin{gather}
\label{Solution_Def}
\begin{aligned}
u =& R,  \\
w =& w(R) \\
&+ \epsilon^2 (b_{1} R_{xx} + b_{2}
R_{x}^{2}) \\
&+ \epsilon^{4} (d_{1} R_{4x} + d_{2}R_{xxx} R_{x}  + d_{3}
R_{xx}^{2} + d_{4} R_{xx} R_{x}^{2} + d_{5} R_{x}^{4}) +
O(\epsilon^{6}),
\end{aligned}
\end{gather}
where the coefficients $b_{i}$ and $d_{i}$ are certain functions of $R$,
and  $\partial_{x}^{n}R = R_{nx}$. Notice that the relation $u=R$ remains undeformed: this can always be achieved modulo the Miura group.  Similarly, the deformed version of Eqs. (\ref{R}) is
\begin{gather}
\label{Riemann_Def}
\begin{aligned}
R_{y} =& \mu R_{x} \\
&+ \epsilon^{2}(\varphi_{1} R_{xxx} +
\varphi_{2} R_{xx} R_{x} + \varphi_{3} R_{x}^3 ) \\
&+ \epsilon^{4} (\rho_{1} R_{5x} + \rho_{2} R_{4x} R_{x}  +
\rho_{3} R_{xxx} R_{xx} + \rho_{4} R_{xx}^{2} R_{x} + \rho_{5}
R_{xxx} R_{x}^{2} + \rho_{6} R_{xx} R_{x}^{3} + \rho_{7}
R_{x}^{5}) \\
&+O(\epsilon^{6}),\\
R_{t} =& (\mu^{2} + R) R_{x} \\
&+\epsilon^{2} (\beta_{1} R_{xxx} + \beta_{2} R_{xx} R_{x}  +
\beta_{3} R_{x}^{2})\\
&+ \epsilon^{4} (\delta_{1} R_{5x} + \delta_{2}R_{4x} R_{x}  +
\delta_{3} R_{xxx} R_{xx}  + \delta_{4} R_{xx}^{2} R_{x} +
\delta_{5} R_{xxx} R_{x}^{2} +
\delta_{6} R_{xx} R_{x}^{3} + \delta_{7} R_{x}^{5} ) \\
&+O(\epsilon^{6}),
\end{aligned}
\end{gather}
where, again,  coefficients are certain functions of $R$
(it can be demonstrated that all odd order corrections in $\epsilon$ must vanish
identically). Substituting Eqs. (\ref{Solution_Def}) into Eqs. (\ref{KP}),
using Eqs. (\ref{Riemann_Def}) and the compatibility condition $R_{yt}
= R_{ty}$, we arrive at the recursive formulae for higher order corrections which are uniquely  expressed in terms of $\mu(R)$ and its derivatives.
Thus, at the order $\epsilon^{2}$ we obtain
\begin{align*}
&b_{1} = \frac{\mu'}{12},& \qquad &b_{2}=\frac{1}{24} (\mu'' -
(\mu')^{3}),&\\
&\varphi_{1} = \frac{\mu'}{12},& \qquad
&\beta_{1} = \frac{1}{12} (2 \mu \mu' + 1),&  \\
& \varphi_{2} = \frac{1}{12} (2 \mu'' - (\mu')^{3}),& \qquad
&\beta_{2} =
\frac{1}{12} (4 \mu \mu'' - 2 \mu (\mu')^{3} + 3 (\mu')^{2}),&  \\
& \varphi_{3} = \frac{1}{24} ( \mu''' - 3 (\mu')^{2} \mu''),&\qquad
&\beta_{3} = \frac{1}{12} (\mu \mu''' + 2 \mu' \mu'' - 3 \mu
(\mu')^{2} \mu'' - (\mu')^{4}),&
\end{align*}
which implies (\ref{R_Def}). Similarly, at the order $\epsilon^{4}$ we have
\begin{align*}
&d_{1} = \frac{1}{720} (3 \mu'' - (\mu')^{3}), \\
&d_{2}=\frac{1}{720} (6 \mu''' - 19 (\mu')^{2} \mu'' +
(\mu')^{5}), \\
&d_{3} = \frac{1}{1440} (9 \mu''' - 16 (\mu')^{2} \mu'' + 4
(\mu')^{5}),\\
& d_{4} = \frac{1}{1440} (11 \mu^{(IV)}- 49
(\mu')^{2} \mu''' - 84 \mu' (\mu'')^{2}  + 52 (\mu')^{4}
\mu''),\\
&d_{5} = \frac{1}{5760} (5 \mu^{(V)}- 32 (\mu')^{2} \mu^{(IV)}-
146 \mu' \mu'' \mu''' + 51 (\mu')^{4} \mu'''   - 44 (\mu'')^{3}  +
216 (\mu')^{3} (\mu'')^{2}).
\end{align*}
The remaining coefficients can be expressed as follows:
\begin{align*}
&\rho_{1} = d_{1},& \qquad &\rho_{2} = d_{1}' + d_{2},& \qquad
&\rho_{3} = d_{2} + 2 d_{3},& \\
&\rho_{4} = d_{3}' + 2 d_{4},& \qquad &\rho_{5} = d_{2}' + d_{4},&
\qquad &\rho_{6} = d_{4}' + 4 d_{5}, \qquad \rho_{7} = d_{5}',&
\end{align*}
\begin{align*}
\delta_{1} &= 2 \mu d_{1} + \frac{1}{144} (\mu')^{2}, \\
\delta_{2} &= 2 \mu (d_{1}' + d_{2}) + \frac{\mu'}{144}  (720
d_{1}- 2 (\mu')^{3}+ 5 \mu''),\\
\delta_{3} &= 2 \mu (d_{2} + 2 d_{3}) + \frac{\mu'}{144}  (1440
d_{1}- 3 (\mu')^{3} + 8 \mu''),\\
\delta_{4} &= 2 \mu (d_{3}' + 2 d_{4})+ \frac{\mu''}{144} (2160
d_{1} + 4 \mu'') + \frac{\mu'}{144} \left (432 d_{2} + 864 d_{3} -
22 (\mu')^{2} \mu'' + 9 \mu''' +
(\mu')^{5}\right),  \\
\delta_{5}&= 2 \mu (d_{2}' + d_{4})+ \frac{\mu''}{48} (480 d_{1} +
\mu'') + \frac{\mu'}{144} \left(720 d_{2} - 16 (\mu')^{2} \mu'' +
7 \mu''' + (\mu')^{5} \right),
 \\
\delta_{6} &= 2 \mu (d_{4}' + 4 d_{5}) + 2 \mu'' (d_{3} + 3 d_{2})
+ \frac{5}{288} \mu''' (d_{1} + 2 \mu'') \\
&+ \frac{\mu'}{288}\left (1440 d_{4} + 18 (\mu')^{4} \mu'' - 78
\mu' (\mu'')^{2} - 35 (\mu')^{2} \mu''' + 11 \mu^{(IV)}
\right), \\
\delta_{7}&= 2 \mu d_{5}' + \mu'' d_{4} + \frac{\mu^{(IV)}}{576}
\left(d_{1} + 2 \mu'' \right) + \frac{\mu'''}{576} (576 d_{2} +
\mu''') \\
&+ \frac{\mu'}{576} \left(2304 d_{5} + 21 (\mu')^{3} (\mu'')^{2} +
6 (\mu')^{4} \mu''' - 48 \mu' \mu'' \mu'''- 8 (\mu')^{2}
\mu^{(IV)}- 24 (\mu'')^{3} + 2 \mu^{(V)}  \right).
\end{align*}
Although we have  calculated dispersive corrections up to the order $\epsilon^6$, the expressions are getting increasingly more complicated. It is important to emphasize that the calculation of dispersive corrections is an entirely {\it algebraic} procedure which does not require  solving differential equations.

\subsection{Two-component reductions}

Two-component reductions of the dKP equation~(\ref{dKP}) are
sought in the form $u = u(R^{1},R^{2})$,
$w = w(R^{1},R^{2})$, where the Riemann invariants $R^{1}$ and $R^{2}$  satisfy the two-component version of Eqs. (\ref{Ri}). One  has $\lambda^i=(\mu^i)^2+u, \ \partial_jw=\mu^j\partial_ju$, where $\mu^j$ and $u$ satisfy
the two-component  Gibbons-Tsarev system (\ref{GT}),
\begin{equation}
\partial_{2} \mu^{1} = \frac{\partial_{2}u}{\mu^{2}-
\mu^{1}}, \qquad \partial_{1} \mu^{2} = \frac{\partial_{1}u}{\mu^{1}-
\mu^{2}}, \qquad \partial_{12} u = \frac{2\; \partial_{1}u\;
\partial_{2} u}{(\mu^{2}- \mu^{1})^{2}}.
\label{2GT}
\end{equation}
We
consider $(1+1)-$dimensional decompositions of the KP equation (\ref{KP})
 obtained by deforming Eqs. (\ref{R}),
\begin{gather}
\label{2Riemann_Deformed}
\begin{aligned}
R^{i}_{y} &= \mu^{i} R^{i}_{x} + \ep^{2}
\left(a^i_jR^j_{xxx}+ b^i_{jk}R^j_{xx}R^k_x+c^i_{jkl}R^j_xR^k_xR^l_x\right)+ O(\ep^{4}),  \\
R^{i}_{t} &= \lambda^{i} R^{i}_{x} + \ep^{2}
\left(A^i_jR^j_{xxx}+ B^i_{jk}R^j_{xx}R^k_x+C^i_{jkl}R^j_xR^k_xR^l_x\right)+ O(\ep^{4}),
\end{aligned}
\end{gather}
where the coefficients $a^i_j, A^i_j$, etc, are certain functions of  $R^1, R^2$; here $i, j, k, l=1,2$.
We point out that  $u(R^{1},R^{2})$ and $w(R^{1},R^{2})$ remain  undeformed (this can always be achieved modulo the Miura group). Moreover, one can show that all terms at the odd powers of $\epsilon$ must vanish identically.
Substituting $u(R^{1},R^{2})$ and $w(R^{1},R^{2})$  into Eqs. (\ref{KP}), using Eqs. (\ref{2Riemann_Deformed}) and imposing the
compatibility condition $R^{i}_{yt} = R^{i}_{ty}$,  one uniquely recovers
 expressions  for the coefficients in terms of $\mu^{i}$ and $u$.
 Although these expressions are quite complicated in general,
 some of them are remarkably simple. Thus,
$$
a^1_1= \frac{1}{12} \frac{\partial_1\mu^1}{\partial_1u}, \qquad
a^1_2=-\frac{1}{12}\frac{\partial_2\mu^2}{\partial_1u}, \qquad
a^2_1=-\frac{1}{12}\frac{\partial_1\mu^1}{\partial_2u}, \qquad
a^2_2=\frac{1}{12}\frac{\partial_2\mu^2}{\partial_2u},
$$
or, in tensor notation,
$a^i_j=\frac{1}{12}(-1)^{i+j}{\partial_j\mu^j}/{\partial_iu}$. Similarly,
$$
A^{1}_{1} =\frac{1}{12} + \frac{\mu^{1} \partial_{1} \mu^{1}}{6
\partial_{1} u},   ~~
A^{1}_{2} = - \frac{(\mu^{1}+\mu^{2}) \partial_{2} \mu^{2}}{12
\partial_{1} u},  ~~
A^{2}_{1} = - \frac{(\mu^{1}+\mu^{2}) \partial_{1} \mu^{1}}{12
\partial_{2} u}, ~~  A^{2}_{2} =\frac{1}{12} +
\frac{\mu^{2} \partial_{2} \mu^{2}}{6
\partial_{2} u},
$$
or $A^i_j=\frac{1}{12}((-1)^{i+j}(\mu^i+\mu^j) \partial_j\mu^j/{\partial_iu}+\delta^i_j).$

\medskip

As a particular case, let us consider
deformations (\ref{2Riemann_Deformed}) such that the
series truncates at the order $\ep^{2}$. This leads to the following two possibilities:

\medskip

\noindent {\bf Case 1.}
\begin{gather}
\label{Trunc1}
\begin{aligned}
\partial_{1}u = (\mu^{2} - \mu^{1}) \partial_{1}
\mu^{1}, \qquad
\partial_{2}u = (\mu^{1} - \mu^{2}) \partial_{2}
\mu^{2}.
\end{aligned}
\end{gather}
Eqs. (\ref{Trunc1}) together with the Gibbons-Tsarev
system~(\ref{2GT}) imply
\begin{equation*}
\mu^{1} + \mu^{2} = \textup{const}.
\end{equation*}
Setting $\mu^1=a+p, \ \mu^2=a-p$ where $a={\rm const}$, and
substituting this representation into the first two equations
(\ref{2GT}), one obtains $u=b-p^2, \ b={\rm const}$.
Then Eq. (\ref{2GT})$_3$ implies
$\partial_{12}(p^3)=0$, so that, up to a reparametrization
of  Riemann invariants, one has
$p=(R^1-R^2)^{1/3}$. Ultimately,
$$
\mu^1=a+(R^1-R^2)^{1/3}, \qquad \mu^2=a+(R^2-R^1)^{1/3}.
$$
Up to a linear transformation of the independent variables, one can set $a=b=0$. This implies $\lambda^i=0$ and $ R^i_t=0$,  so that the system $(\ref{2Riemann_Deformed})_2$ becomes trivial, while the system $(\ref{2Riemann_Deformed})_1$ reduces to the stationary Boussinesq reduction of the dKP equation, $(u u_{x} + \frac{\epsilon^{2}}{12} u_{xxx} )_x+u_{yy}=0$.

\medskip

\noindent {\bf Case 2.}
\begin{gather}
\label{Trunc2}
\begin{aligned}
\partial_{1}u = \frac{1}{3}(\mu^{1} - \mu^{2}) \partial_{1}
\mu^{1}, \qquad
\partial_{2}u = \frac{1}{3} (\mu^{2} - \mu^{1}) \partial_{2}
\mu^{2}.
\end{aligned}
\end{gather}
Eqs. (\ref{Trunc2}) together with the Gibbons-Tsarev
system~(\ref{2GT}) imply that, up to a reparametrization of Riemann invariants, one can set
\begin{equation*}
\mu^{1} = \frac{3}{4}R^{1} +\frac{1}{4}R^{2}, \qquad \mu^{2} = \frac{3}{4} R^{2} +\frac{1}{4}R^{1}, \qquad u = \frac{1}{16}(R^{1}-R^{2})^2.
\end{equation*}
The corresponding $\epsilon^{2}-$coefficients in Eqs.
(\ref{2Riemann_Deformed}) take the form
$$
a^{1}_{1} = a^2_1= \frac{1}{2 (R^{1}-R^{2})},  \qquad a^{1}_{2} = a^2_2=
\frac{1}{2 (R^{2}-R^{1})},
$$
$$
A^{1}_{1} = \frac{5 R^{1} + R^{2}}{6 (R^{1}-R^{2})}, \qquad A^{1}_{2} =  \frac{R^{1}+R^{2}}{2 (R^{2}-R^{1})},  \qquad A^{2}_{1} = \frac{R^{1}+R^{2}}{2 (R^{1}-R^{2})},   \qquad A^{2}_{2} =  \frac{R^{1} + 5 R^{2}}{6 (R^{2}-R^{1})},
$$
etc. Introducing the new dependent variables $v=(R^1+R^2)/2$ and $ u =
(R^{1}-R^{2})^2/16$, one can rewrite Eqs.
(\ref{2Riemann_Deformed}) as
\begin{gather*}
\begin{aligned}
&u_{y} = (u v)_{x}, \\
&v_{y} = v v_{x} + u_{x} + \frac{\epsilon^{2}}{4}
\left(\frac{u_{xx}}{u} - \frac{1}{2} \frac{u_{x}^{2}}{u^{2}}
\right)_{x},
\end{aligned}
\end{gather*}
and
\begin{gather*}
\begin{aligned}
&u_{t} = (2 u + v^{2}) u_{x} + 2 u v v_{x} +
\frac{\epsilon^{2}}{12} \left(4 u_{xx} - 3 \frac{u_{x}^{2}}{u}
\right)_{x}, \\
&v_{t} = 2 v u_{x} + (2 u + v^{2}) v_{x} + \frac{\epsilon^{2}}{2}
\left[v \left(\frac{u_{xx}}{u} -\frac{1}{2}
\frac{u_{x}^{2}}{u^{2}} \right) + \frac{1}{3} \left(2 v_{xx} + 3
\frac{u_{x}}{x} v_{x} \right) \right]_{x},
\end{aligned}
\end{gather*}
respectively. This is the well-known Zakharov reduction of KP to NLS \cite{Z}.

\subsection{Waterbag reduction}
It was observed in \cite{Pavlov} that  $n$-component reductions of  dKP can also be sought in
the form $u = u(v^{1}, v^{2},\dots, v^{n})$, $w =
w(v^{1}, v^{2},\dots, v^{n})$, where the fields $v^{i}$ satisfy the
equations
\begin{equation}
\label{WBtype} v^{i}_{y} = \left(\frac{(v^{i})^{2}}{2} + u
\right)_{x}, \qquad v^{i}_{t} = \left(\frac{(v^{i})^{3}}{3} + u
v^{i} + w \right)_{x}.
\end{equation}
The system~(\ref{WBtype}) is automatically compatible provided  $u$ and $w$ solve Eqs. (\ref{dKP}).
The substitution into  Eq. (\ref{dKP})$_{2}$ implies
\begin{equation}
\label{W_Trunc}
\partial_{i}w = \left(\sum_{k=1}^{n} \partial_{k} u + v^{i}
\right)\partial_{i} u,
\end{equation}
here $\partial_i=\partial_{v^i}$. It turns out that Eq. (\ref{dKP})$_{1}$ is satisfied identically
modulo Eq. (\ref{W_Trunc}). The consistency conditions
$\partial_{i} \partial_{j}w = \partial_{j} \partial_{i}w$ imply  \cite{Pavlov}
\begin{equation}
\label{u_Trunc_constr}
\partial_{ij}u = \frac{\left(\sum_{k \neq j} \partial_{ki}u \right)
\partial_{j}u + \left(\sum_{k \neq i} \partial_{kj}u \right)
\partial_{i}u }{\partial_{i}u - \partial_{j}u + v^{i}-v^{j}},
\end{equation}
which can be viewed as an analogue of the Gibbons-Tsarev system (\ref{GT}). In  the two-component case, $n=2$, the
system~(\ref{W_Trunc}) simplifies to
\begin{equation}
\label{W_Trunc_red2}
\partial_{1}w = (\partial_{1}u + \partial_{2} u + v^{1})
\partial_{1} u, \qquad \partial_{2}w = (\partial_{1}u +
\partial_{2} u + v^{2})
\partial_{2} u,
\end{equation}
and the compatibility condition $\partial_{1}\partial_{2} w =
\partial_{2}\partial_{1} w$ takes the form
\begin{equation}
\partial_{12}u = \frac{\partial_{2}u\; \partial_{11}u - \partial_{1}u
\;
\partial_{22}u}{\partial_{1}u -\partial_{2}u + v^{1}-v^{2}};
\end{equation}
this is an analogue of Eqs. (\ref{2GT}). The special case
\begin{equation}
\label{Wb} u = \alpha _{1} v^{1} +\alpha_{2} v^{2}, \qquad w = \frac{\alpha_{1}
(v^{1})^{2} + \alpha_{2} (v^{2})^{2}}{2} + (\alpha_{1} + \alpha_{2}) (\alpha_{1} v^{1}
+ \alpha_{2} v^{2}),
\end{equation}
where $\alpha_{1}$ and $\alpha_{2}$ are constants, is known as the {\em
waterbag} reduction \cite{Kodama}. We seek a deformation of the two-component waterbag
reduction (\ref{WBtype})  in the form
\begin{gather}
\label{WbDeform}
\begin{aligned}
v^{i}_{y} &= \left(\frac{(v^{i})^{2}}{2} + u \right)_{x} +
\epsilon^{2} P^i
+ O(\epsilon^{4}), \\
v^{i}_{t} &= \left(\frac{(v^{i})^{3}}{3} + u v^{i} + w \right)_{x}
+ \epsilon^{2} Q^i + O(\epsilon^{4}),
\end{aligned}
\end{gather}
$i=1, 2$. Modulo the Miura group, we assume that Eqs. (\ref{Wb}) remain undeformed. Substituting Eqs. (\ref{Wb}) and (\ref{WbDeform}) into (\ref{dKP}) and using  the
compatibility conditions $v^{i}_{yt} = v^{i}_{ty}$,   we
obtain explicit expressions for $P^i$ and $Q^i$. Thus,  Eqs. $(\ref{WbDeform})_1$ take the form
\begin{gather}
\label{P11}
\begin{aligned}
v^{1}_{y} &= \left(\frac{(v^{1})^{2}}{2} + u \right)_{x} +
\frac{\epsilon^{2}}{\alpha_1} P
+ O(\epsilon^{4}), \\
v^{2}_{y} &= \left(\frac{(v^{2})^{2}}{2} + u \right)_{x} -
\frac{\epsilon^{2}}{\alpha_2} P
+ O(\epsilon^{4}),
\end{aligned}
\end{gather}
where
\begin{align*}
P&= \frac{\Delta + 2 \alpha_{1} - \alpha_{2}}{12 \Delta}v^1_{xxx} - \frac{\Delta +
\alpha_{1} - 2 \alpha_{2}}{12  \Delta}v^2_{xxx} \\
\ \\
&+ \frac{m- \alpha_{2}\Delta (\Delta - \alpha_{1} - 2 \alpha_{2})
}{12 \alpha_{1} \alpha_{2} \Delta^{2}}v^1_{xx}v^1_x - \frac{n
\Delta + m } {12 \alpha_{1} \alpha_{2}
\Delta^{2}}(v^1_{xx}v^2_x+v^2_{xx}v^1_x) +
\frac{m+\alpha_{1}\Delta (\Delta + 2 \alpha_{1} + \alpha_{2}) }{12
\alpha_{1} \alpha_{2}
\Delta^{2}}v^2_{xx}v^2_x\\
\ \\
& - \frac{a_{2} \Delta^{2} +m}{24 \alpha_{1} \alpha_{2}
\Delta^{3}}(v^1_x)^3
+ \frac{\alpha_{2} \Delta^{2} + 3 m}{24 \alpha_{1} \alpha_{2}
\Delta^{3}}(v^1_x)^2v^2_x+
 \frac{\alpha_{1} \Delta^{2} - 3
m}{24 \alpha_{1} \alpha_{2} \Delta^{3}}v^1_x(v^2_x)^2 -\frac{\alpha_{1} \Delta^{2} - m}{24
\alpha_{1} \alpha_{2}
\Delta^{3}}(v^2_x)^3;
\end{align*}
here
$$
\Delta = v^{1} - v^{2}, \qquad m = \alpha_{1}^{2} (\alpha_{1} - 2
\alpha_{2})-
 \alpha_{2}^{2} (\alpha_{2} - 2 \alpha_{1}), \qquad n = \alpha_{1}^{2} + \alpha_{1}
\alpha_{2}+ \alpha_{2}^{2}.
$$
Notice that, although  the hydrodynamic part of Eqs. (\ref{P11}) is conservative, the expression $P$ is not a total $x$-derivative. The expressions for $Q^i$ are not presented here due to their complexity.

\section{Deformations of the dTl reductions}

Expanding the r.h.s. in Eqs. (\ref{TL}) one obtains
\begin{gather}
\label{TL_exp}
\begin{aligned}
u_{y}/u &= w_{x} - \frac{\epsilon}{2} w_{xx} +
\frac{\epsilon^{2}}{6} w_{xxx} + \dots + (-1)^{n+1}
\frac{\epsilon^{n}}{n!} w_{nx} + \dots, \\
w_{t} &= u_{x} + \frac{\ep}{2} u_{xx} + \frac{\ep^{2}}{6} u_{xxx}+ \dots
+ \frac{\ep^{n}}{n!} u_{nx} + \dots
\end{aligned}
\end{gather}
The corresponding dispersionless limit, the dTl equation, is
\begin{equation}
\label{dTL}
u_{y} = u w_{x}, \qquad w_{t} = u_{x}.
\end{equation}
It admits  one-component reductions of the form
\begin{equation*}
u = R, \qquad w = w(R),
\end{equation*}
where $w'=1/\mu$, and $R(x, y, t)$ satisfies a pair of Hopf-type equations
\begin{equation}
\label{dTL_Red} R_{t} = \mu R_{x}, \qquad R_{y} =
\frac{R}{\mu} R_{x};
\end{equation}
here $\mu(R)$ is an arbitrary function. Deformations are sought in the form
\begin{gather}
\label{SolutionTL_Def}
\begin{aligned}
u =& R, \\
w =& w(R)
 + \epsilon a R_{x}
+ \epsilon^2 (b_{1} R_{xx} + b_{2} R_{x}^{2}) + \epsilon^{3}
(c_{1} R_{xxx} + c_{2} R_{xx} R_{x}  + c_{3} R_{x}^{3})+
O(\epsilon^{4}),
\end{aligned}
\end{gather}
where $a$, $b_{i}$, $c_{i}$ are certain functions of $R$, and
\begin{gather}
\label{RiemannTL_Def}
\begin{aligned}
R_{t} =& \mu R_{x} +\epsilon^{2} (\beta_{1} R_{xxx} + \beta_{2}
R_{xx} R_{x}  +
\beta_{3} R_{x}^{3}) + O(\ep^{4}),\\
R_{y} =& \frac{R}{\mu} R_{x}+ \epsilon^{2}(\varphi_{1} R_{xxx} +
\varphi_{2} R_{xx} R_{x} + \varphi_{3} R_{x}^3 ) +
O(\epsilon^{4}).
\end{aligned}
\end{gather}
We point out that the relation $u=R$ remains undeformed (modulo the Miura group), and that all odd order corrections in $\epsilon$ in the equations~(\ref{RiemannTL_Def})
vanish identically. Substituting (\ref{SolutionTL_Def})
into~(\ref{TL_exp}), using (\ref{RiemannTL_Def}) and the
compatibility condition $R_{yt} = R_{ty}$,  one can recursively calculate all higher order corrections in terms of
$\mu$ and its derivatives at  different powers of $\epsilon$:\\
Order $\ep$:
\begin{equation*}
a = \frac{1}{2 \mu};
\end{equation*}
Order $\ep^{2}:$
\begin{gather*}
\begin{aligned}
b_{1} = &-\frac{R \mu'- 2 \mu }{12 \mu^{2}}, \\
 b_{2} =&-\frac{R
\mu^{2} \mu'' - R^{2} (\mu')^{3} + 3 \mu^{2} \mu'}{24 \mu^{4}},\\
\beta_{1} =& \frac{R \mu'}{12}, \\
 \beta_{2} = &\frac{R}{12
\mu^{2}} \left((\mu')^{2} (\mu - R \mu') + 2 \mu^{2} \mu'' \right),\\
\beta_{3} =& \frac{R}{24 \mu^{3}} \left(2 R (\mu')^{4} + 2 \mu^{2}
\mu' \mu'' - \mu (\mu')^{2} (2 \mu' + 3 R \mu'') + \mu^{3} \mu'''
\right),\\
\vphi_{1} =& - \frac{R}{12 \mu^{2}} (R \mu' - \mu ), \\
\vphi_{2} = &\frac{R}{12 \mu^{4}} \left(2 R \mu (\mu')^{2} + R^{2}
(\mu')^{3} - \mu^{2} (3 \mu' + 2 R \mu'') \right),\\
\vphi_{3} =& - \frac{R}{24 \mu^{5}} \left(4 R^{2} (\mu')^{4} - 2
\mu^{2} \mu' (R \mu'' + \mu') - R \mu (\mu')^{2} (3 R \mu'' + 2
\mu') + \mu^{3} (R \mu''' + 2 \mu'') \right);
\end{aligned}
\end{gather*}
Order $\ep^{3}$:
\begin{gather*}
\begin{aligned}
c_{1} =& -\frac{R \mu' -\mu}{24 \mu^{2}}, \\
 c_{2} = &\frac{
R^{2} (\mu')^{3}+ 2 R \mu (\mu')^{2}  - \mu^{2} (2 R \mu'' + 3
\mu')}{24 \mu^{4}}\\
c_{3} =& - \frac{1}{48 \mu^{5}} \left(4 R^{2} (\mu')^{4} - 2
\mu^{2} \mu' (R \mu'' + \mu') - R \mu (\mu')^{2} (3 R \mu'' + 2
\mu') + \mu^{3} (R \mu''' + 2 \mu'') \right),
\end{aligned}
\end{gather*}
etc. Again, this calculation is an entirely algebraic procedure.

\section{Recostruction of dispersive terms in 2D}

Given a dispersionless integrable system in $2+1$ dimensions, how can one reconstruct the corresponding dispersive counterpart? This natural problem was  first addressed by Zakharov in \cite{Zakharov}, based on the quantization of the corresponding dispersionless Lax pair. This, however, does not work  when the Lax pair is `sufficiently complicated'. We propose an alternative approach to this problem, based on the  requirement that {\it all} hydrodynamic reductions of the dispersionless system should be inherited by its dispersive counterpart. This  imposes strong restrictions on the structure of dispersive terms. The following simple examples illustrate the method; further results in this direction will be reported elsewhere.

\subsection{Generalized KP equation}
A two-component form of the generalized KP equation (\ref{kp_Ext}) is
\begin{gather}
\label{dKP_Ext}
\begin{aligned}
 u_{t} - u u_{x} + \epsilon (A_{1} u_{xx} + A_{2} u_{x}^{2})
 + \epsilon^{2} (B_{1} u_{xxx} + B_{2} u_{x} u_{xx} + B_{3} u_{x}^{3})
&= w_{y},\\
u_{y} &= w_{x},
\end{aligned}
\end{gather}
where $A_{i}$ and $B_{i}$ are certain  functions of $u$. Let us require that {\it any} one-component reduction (\ref{R}) of its dispersionless limit can be deformed, in the form (\ref{Riemann_Def}),  into a reduction of Eqs. (\ref{dKP_Ext}). Thus, whenever one obtains a relation involving $\mu(R)$ and its derivatives, all coefficients must  be set equal to zero: this ensures that $\mu(R)$ remains arbitrary. Looking at different powers of $\epsilon$ we arrive at the
following results: \\
\noindent order $\epsilon$\;:
\begin{equation*}
A_{1} = 0, \qquad A_{2} = 0;
\end{equation*}
\noindent order $\epsilon^{2}$:
\begin{equation*}
B_{1} = \textup{const}, \qquad B_{3} = \frac{B_{2}'}{2};
\end{equation*}

\noindent order $\epsilon^{4}$:
\begin{equation*}
3 B_{1} B_{2}' - 2 B_{2}^{2} = 0.
\end{equation*}
This implies  either $B_{2} = 0$, or $B_{2} = -3
B_{1}/(2 u + c_{0})$, where $c_{0}$ is an arbitrary integration
constant. It follows that only the case
\begin{equation*}
B_{2} = 0.
\end{equation*}
is admissible. Thus,  the KP equation is the only  possible extension,
within the class (\ref{dKP_Ext}), which inherits deformations of
all hydrodynamic reductions.

\subsection{BKP and CKP equations}

Let us consider  equations of the form
\begin{gather}
\label{KK}
\begin{aligned}
u_{t} - 5 (u^{2} + w) u_{x} - 5 u w_{x} + 5 w_{y} + \epsilon^{2}
(A_{0} u u_{xxx} + A_{1} w_{xxx} + A_{2} u_{xxx}) + \epsilon^{4}
A_3 \; u_{xxxxx}&=0, \\
u_{y}& = w_{x},
\end{aligned}
\end{gather}
where $A_0 - A_3$ are arbitrary constants. Deformed one-component reductions are sought in the form
\begin{gather}
\label{KKsol}
\begin{aligned}
u =& R,  \\
w =& w(R)  \\
&+ \epsilon^2 (b_{1} R_{xx} + b_{2}
R_{x}^{2}) \\
&+ \epsilon^{4} (d_{1} R_{4x} + d_{2}R_{xxx} R_{x}  + d_{3}
R_{xx}^{2} + d_{4} R_{xx} R_{x}^{2} + d_{5} R_{x}^{4}) +
O(\epsilon^{6}),
\end{aligned}
\end{gather}
where  $R$ obeys the system of the form
\begin{gather}
\label{KKred}
\begin{aligned}
R_{y} =& \mu R_{x} \\
&+ \epsilon^{2}(\varphi_{1} R_{xxx} +
\varphi_{2} R_{xx} R_{x} + \varphi_{3} R_{x}^3 ) \\
&+ \epsilon^{4} (\rho_{1} R_{5x} + \rho_{2} R_{4x} R_{x}  +
\rho_{3} R_{xxx} R_{xx} + \rho_{4} R_{xx}^{2} R_{x} + \rho_{5}
R_{xxx} R_{x}^{2} + \rho_{6} R_{xx} R_{x}^{3} + \rho_{7}
R_{x}^{5}) \\
&+O(\epsilon^{6}),\\
R_{t} =& 5 (R^{2} + w + \mu (R-\mu) ) R_{x} \\
&+\epsilon^{2} (\beta_{1} R_{xxx} + \beta_{2} R_{xx} R_{x}  +
\beta_{3} R_{x}^{2})\\
&+ \epsilon^{4} (\delta_{1} R_{5x} + \delta_{2}R_{4x} R_{x}  +
\delta_{3} R_{xxx} R_{xx}  + \delta_{4} R_{xx}^{2} R_{x} +
\delta_{5} R_{xxx} R_{x}^{2} +
\delta_{6} R_{xx} R_{x}^{3} + \delta_{7} R_{x}^{5} ) \\
&+O(\epsilon^{6});
\end{aligned}
\end{gather}
here $\mu(R)$ is an arbitrary function, and $w'=\mu$. Substituting Eqs. (\ref{KKsol}) and~(\ref{KKred}) into Eqs. (\ref{KK}), we obtain
$\varphi$, $\rho$, $\beta$, $\delta$  in terms of $b$, $d$,
etc. The compatibility condition $R_{ty} = R_{yt}$ provides the
following constraints on the constants $A_{i}$'s, \\

order $\epsilon^{2}:$
\begin{equation*}
A_{1} = A_{0},
\end{equation*}
\indent order $\epsilon^{4}:$
\begin{align*}
&A_3 = - \frac{A_{0}^{2}}{25}, \qquad
5 A_{0}^{2} - 7 A_{0} A_{2} + 2 A_{2}^{2} = 0.
\end{align*}
The quadratic equation leads to either
$A_{2} = A_{0}$,
or
$ A_{2} = \frac{5}{2} A_{0},$
which correspond  the
BKP  and CKP equations, respectively. We have verified that, up to the order
$\epsilon^{6}$, no extra constraints appear,  and that all  coefficients in the
expansions~(\ref{KKsol}) and~(\ref{KKred}) are certain expressions
in terms of  $\mu$. This calculation shows that the reconstruction of dispersive corrections is essentially non-unique.

\subsection{The `universal'  equation}

In this section we study dispersive deformations of the so-called `universal' equation,
\begin{gather}
\label{uni}
\begin{aligned}
u_{y} =& w_{x}, \\
u_{t} =& w_{y}  + u w_{x} - w u_{x},
\end{aligned}
\end{gather}
which was discussed in a series of publications \cite{Pav, Shabat, Dun}, etc. The structure of hydrodynamic reductions in this case is particularly simple, namely, $n$-phase solutions are given by the formulae
$$
u=R^1+...+R^n, ~~~~ w=f^1(R^1)+...+f^n(R^n)-\frac{1}{2}(R^1+...+R^n)^2,
$$
where $f^i(R^i)$ are n arbitrary functions of one variable, and the phases $R^i$ satisfy Eqs. (\ref{Ri}) with
$$
\mu^i={f^i}'-(R^1+...+R^n), ~~~~ \lambda^i=(\mu^i)^2+u\mu^i-w;
$$
notice that $\mu^i=\partial_iw$. In the one-component case one has $u=R, \ w=w(R)$, where the single phase $R$ satisfies the equations
\begin{equation}
 R_y=\mu(R) R_x,  ~~~~ R_t=(\mu^2+R\mu-w) R_x,
\label{Rii}
\end{equation}
$w'=\mu$.
We have found   the following dispersive deformation of Eq. (\ref{uni}),
\begin{gather}
\label{Univ_Def0}
\begin{aligned}
u_{y} =& w_{x}, \\
u_{t} =& w_{y}  + u w_{x} - w u_{x}+\tau \ep^{4} (u_{xx} w_{xxx} - w_{xx} u_{xxx})+O(\ep^{8}),
\end{aligned}
\end{gather}
which inherits deformations of  all one-components reductions (\ref{Rii}) up to the order $\ep^8$. Our procedure, however, does not work beyond this order, leading to the trivial condition $\tau=0$.
We would like to emphasize that the  extension of  two-component reductions leads to the condition $\tau=0$ already at the order $\ep^4$. The deformation (\ref{Univ_Def0}) is found by seeking dispersive extensions of Eq. (\ref{uni}) in the form
\begin{gather}
\label{Univ_Def}
\begin{aligned}
u_{y} =& w_{x}, \\
u_{t} =& w_{y}  + u w_{x} - w u_{x}\\
 &+ A \ep (u w_{xx} - w
u_{xx})\\
&+ B_1 \ep^{2} (u w_{xxx} - w u_{xxx}) + B_2 \ep^{2} (u_{x} w_{xx}
-
w_{x} u_{xx})\\
&+ C_1 \ep^{3} (u w_{4x} - w u_{4x}) + C_2 \ep^{3} (u_{x} w_{xxx}
- w_{x} u_{xxx}) \\
&+ D_1 \ep^{4} (u w_{5x} - w u_{5x}) + D_2 \ep^{4} (u_{x} w_{4x} -
w_{x} u_{4x}) + D_3 \ep^{4} (u_{xx} w_{xxx} - w_{xx} u_{xxx}) \\
&+ E_1 \ep^{5} (u w_{6x} - w u_{6x}) + E_2 \ep^{5} (u_{x} w_{5x} -
w_{x} u_{5x}) + E_3 \ep^{5} (u_{xx} w_{4x} - w_{xx} u_{4x}) + \dots,
\end{aligned}
\end{gather}
where the coefficients $A$, $B_{i}$, etc. are constants. Notice that one can simplify Eq. (\ref{Univ_Def}) by using Miura-type transformations of the form
$$
u\to u+\alpha \epsilon u_x+ \beta \epsilon^2 u_{xx}+..., ~~~  w\to w+\alpha \epsilon w_x+ \beta \epsilon^2 w_{xx}+...,
$$
where $\alpha, \beta, ...$ are arbitrary constants. In particular,
using this freedom one can set  $B_2=C_2=D_2=E_2=0$, etc.
In what follows, we assume this normalization.
Let us require that any one-component reduction (\ref{Rii})
can be extended as
\begin{gather}
\label{A1}
\begin{aligned}
u =& R,  \\
w =& w(R) \\
&+ \ep a R_{x}+ \epsilon^2 (b_{1} R_{xx} + b_{2} R_{x}^{2})\\
&+
\ep^{3} (c_1 R_{xxx} + c_2 R_{x} R_{xx} + c_3 R_{x}^{3})\\
&+\ep^{4} \left(d_{1} R_{4x} + d_{2} R_{3x} R_{x} + d_{3}
R_{xx}^{2} +d_{4} R_{xx} R_{x}^{2} + d_{5} R_{x}^{4} \right)+ \dots,
\end{aligned}
\end{gather}
where $R$ satisfies the system
\begin{gather}
\label{A2}
\begin{aligned}
R_{y} =& \mu R_{x} \\
&+ \ep (\sigma_{1} R_{xx} + \sigma_{2} R_{x}^{2}) \\
&+ \epsilon^{2}(\varphi_{1} R_{xxx} + \varphi_{2} R_{xx} R_{x} +
\varphi_{3} R_{x}^3 )\\
&+ \ep^{3} (\psi_{1} R_{4x} + \psi_{2} R_{x} R_{xxx} + \psi_{3}
R_{xx}^{2}+
\psi_{4} R_{xx} R_{x}^{2} + \psi_{5} R_{x}^{4})\\
&+ \ep^{4} \left(\rho_{1} R_{5x} + \rho_{2} R_{4x} R_{x} +
\rho_{3} R_{3x} R_{xx} + \rho_{4} R_{xx}^{2} R_{x} +
\rho_{5} R_{3x} R_{x}^{2} + \rho_{6} R_{xx} R_{x}^{3}
+ \rho_{7} R_{x}^{5}  \right)\\
&+ \dots, \\
R_{t} =& ( \mu^{2}+R \mu - w ) R_{x}\\
 &+ \ep (\alpha_{1}
R_{xx}
+ \alpha_{2} R_{x}^{2}) \\
&+\epsilon^{2} (\beta_{1} R_{xxx} + \beta_{2} R_{xx} R_{x}  +
\beta_{3} R_{x}^{2})\\
&+ \ep^{3} (\gamma_{1} R_{4x} + \gamma_{2} R_{x} R_{xxx} +
\gamma_{3} R_{xx}^{2}+ \gamma_{4} R_{xx} R_{x}^{2} + \gamma_{5}
R_{x}^{4})+
\psi_{4} R_{xx} R_{x}^{2} + \psi_{5} R_{x}^{4})\\
&+ \ep^{4} \left(\delta_{1} R_{5x} + \delta_{2} R_{4x} R_{x} +
\delta_{3} R_{3x} R_{xx} + \delta_{4} R_{xx}^{2} R_{x} +
\delta_{5} R_{3x} R_{x}^{2} + \delta_{6} R_{xx} R_{x}^{3} +
\delta_{7} R_{x}^{5} \right) \\
&+ \dots,
\end{aligned}
\end{gather}
$w'=\mu$. Substituting (\ref{A1}) into (\ref{Univ_Def}), using Eqs. (\ref{A2}) and the compatibility conditions $R_{yt}=R_{ty}$ (where $\mu$ is assumed to be {\it arbitrary}), at different orders in $\ep$ we get the following conditions for
the coefficients of~(\ref{Univ_Def}),
 \\
order $\ep$:
$$
A = 0,
$$
order $\ep^{2}$:
$$
B_{1} = 0,
$$
order $\ep^{3}$:
$$
C_{1} = 0,
$$
order $\ep^{4}$:
$$
D_{1} = 0,
$$
order $\ep^{5}$:
$$
E_{1} = E_{3}=0,
$$
etc. Notice that there is no condition on the coefficient $D_3=\tau$. This results in Eq. (\ref{Univ_Def0}). The corresponding Eqs. (\ref{A1}), (\ref{A2}) take the form
\begin{gather}
\begin{align*}
u =& R,  \\
w =& w(R) \\
&+\ep^{4} \left(d_{1} R_{4x} + d_{2} R_{3x} R_{x} + d_{3}
R_{xx}^{2} +d_{4} R_{xx} R_{x}^{2} + d_{5} R_{x}^{4} \right)+ O(\ep^{8}),
\end{align*}
\end{gather}
and

\begin{gather}
\begin{align*}
R_{y} =& \mu R_{x} \\
&+ \ep^{4} \left(\rho_{1} R_{5x} + \rho_{2} R_{4x} R_{x} +
\rho_{3} R_{3x} R_{xx} + \rho_{4} R_{xx}^{2} R_{x} +
\rho_{5} R_{3x} R_{x}^{2} + \rho_{6} R_{xx} R_{x}^{3}
+ \rho_{7} R_{x}^{5}  \right)\\
&+ O(\ep^{8}), \\
R_{t} =& ( \mu^{2}+R \mu - w ) R_{x}\\
&+ \ep^{4} \left(\delta_{1} R_{5x} + \delta_{2} R_{4x} R_{x} +
\delta_{3} R_{3x} R_{xx} + \delta_{4} R_{xx}^{2} R_{x} +
\delta_{5} R_{3x} R_{x}^{2} + \delta_{6} R_{xx} R_{x}^{3} +
\delta_{7} R_{x}^{5} \right) \\
&+ O(\ep^{8}),
\end{align*}
\end{gather}
respectively. Here all coefficients are certain explicit expressions in terms of $\mu$ and its derivatives,
\begin{align*}
d_{1} &= 0, \qquad d_{2} = \frac{\tau \mu''}{3 \mu'' - 2}, \qquad
d_{3} = - \frac{3 \tau \mu''}{3 \mu'' - 2}, \qquad d_{4} =
\frac{\tau (3 \mu'' + 4) \mu'''}{3 (\mu'')^{2} - 11 \mu'' + 6},\\
d_{5} &= \frac{\tau \left((74 - 19 \mu'' - 12 (\mu'')^{2})
(\mu''')^{2} + (6 - 23 \mu'' - 17 (\mu'')^{2} + 8 (\mu'')^{3})
\mu^{(IV)} \right)}{8 (\mu''-3)^{2} (2 - 5 \mu'' + 3 (\mu'')^{2})},
\end{align*}
\begin{align*}
&\delta_{1} = 0, \qquad \delta_{2} = (R + 2 \mu') \rho_{2}, \qquad
\delta_{3} = (R + 2 \mu') \rho_{3}, \\
&\delta_{4} = (R + 2 \mu') \rho_{4} + 3 \mu'' \rho_{3} + 3 \tau
\mu'' - d_{3}, \\
&\delta_{5} = (R + 2 \mu') \rho_{5} + 5 \mu'' d_{2} - \tau \mu'' -
d_{2}, \\
&\delta_{6} = (R+2\mu') \rho_{6} + 2  \mu''' (3 d_{2} + d_{3}) +
\tau \mu''' + 5 \mu'' d_{4} - d_{4}, \\
&\delta_{7} = (R + 2 \mu') \rho_{7} + \mu^{(IV)} d_{2} + \mu'''
d_{4} + 4 \mu'' d_{5} - d_{5},
\end{align*}
\begin{align*}
&\rho_{1} =0, \qquad &\rho_{2} = d_{2}, \qquad  &\rho_{3} = d_{2}
+3d_{3},\\
&\rho_{4} = d_{3}' + 2 d_{4}, \qquad  &\rho_{5}= d_{2}' + d_{4},
\qquad &\rho_{6} = d_{4}' + 4 d_{5}, &\rho_{7} = d_{5}'.
\end{align*}

\medskip

\noindent{\bf Remark.}
The paper \cite{Dun} provides a  multi-parameter deformation of the universal hierarchy with  the first few terms given by
\begin{align*}
u_{y} =& w_{x}, \\
u_{t} =& w_{y} + u w_{x} - w u_{x} - \frac{\ep^{2} \mu^{2}}{2
\lambda^{4}} \left(u_{xx} w_{xxx} - w_{xx} u_{xxx} \right)+ \frac{\ep^{3} \mu^{2}}{2 \lambda^{5}} \left(u_{xx} w_{4x} -
w_{xx} u_{4x} \right) \\
& + \frac{\ep^{3} \mu^{3}}{3 \lambda^{6}}
\left(u_{xxx} w_{4x} - w_{xxx} u_{4x} \right) - \frac{\ep^{4}
\mu^{2}}{2 \lambda^{6}} \left(u_{xx} w_{5x} - w_{xx} u_{5x}
\right)+ \dots,
\end{align*}
Notice that this deformation belongs to the class (\ref{Univ_Def}). According to our calculations, it {\it does not} inherit hydrodynamic reductions of the universal equation.

\section{Concluding remarks}

Several natural questions arise in the present context:

\medskip

\noindent {\bf 1.}
The results of \cite{Baikov, Dub1, Zhang} suggest that any evolution system in $1+1$ dimensions is quasitrivial, that is,  transformable to its dispersionless limit  by a quasi-Miura transformation. In particular,  deformed hydrodynamic reductions constructed in this paper  are quasitrivial as well. It would be of interest to extend these quasitriviality type results to the corresponding integrable systems in $2+1$ dimensions such as the full KP and Toda lattice equations.

\medskip

\noindent {\bf 2.}
As demonstrated in \cite{Zakharov}, dispersionless limits of integrable systems in $2+1$ dimensions come with  Lax pairs of the form
\begin{equation}
\psi_y=A({\bf u}, \psi_x), ~~~ \psi_t=B({\bf u}, \psi_x),
\label{Lax}
\end{equation}
which generate  equations for the fields ${\bf u}(x, y, t) $ via the compatibilty conditions
$$
A_t-B_y+\{A, B\}=0;
$$
here $\{A, B\}=A_{\xi}B_x-B_{\xi}A_x$ denotes the standard Poisson bracket of $A$ and $B$, and $\xi=\psi_x$. It was shown recently (see e.g. \cite{Fer5, Fer7, Moro}) that, conversely, for broad classes of multi-dimensional dispersionless systems, the existence of Lax pairs of the form (\ref{Lax}) is necessary and sufficient for the integrability. In general, the dependence of $A$ and $B$ on $\xi$ can be rather non-trivial,  which makes a direct `quantization'  of such Lax pairs difficult. Several ways to solve the problem of quantization of  dispersionless Lax pairs of the form (\ref{Lax}) were proposed in the literature, in particular, including the Moyal deformation of the Poisson bracket $\{\ , \ \}$, see e.g. \cite{Zakharov, Strachan2, Sz, Bla2, Dun}. These methods work well for a limited number of  examples, allowing one to reconstruct the admissible dispersive terms, however, they meet difficulties when applied to Lax pairs with a more complicated dependence on $\xi$. Our approach to the problem of quantization is to allow only those dispersive corrections for which {\it all} hydrodynamic reductions of the dispersionless system can be deformed into reductions of its dispersive counterpart. We have demonstrated that this requirement is very restrictive indeed,  and imposes strong constraints on the structure of possible dispersive terms. It would be challenging to apply this recipe to other classes of dispersionless integrable systems obtained in the literature, and to classify the associated soliton systems in $2+1$ dimensions.

\medskip

\noindent {\bf 3.} Our approach provides an infinity of  decompositions of  a given $(2+1)$-dimensional integrable soliton equation into a pair of commuting $(1+1)$-dimensional flows, which are parametrized
by arbitrarily many functions of one variable. An alternative construction is known  as the method of  symmetry constraints, or potential-eigenfunction constraints, see e.g. \cite{Kon1, Kon2, Kon3} and references therein. It would be of interest to understand whether generic deformed hydrodynamic reductions possess any kind of  `symmetry'  interpretation.

\medskip

\noindent {\bf 4.} Solutions of dispersionless systems coming from hydrodynamic reductions are known to break down in finite time. The addition of  dispersive corrections can be viewed as a regularization procedure preventing the gradient catastrophe. Although the structure of higher order corrections can be rather complicated in general, it is the very first term that seems to be of prime importance. Thus, our procedure gives a canonical way to regularize hydrodynamic reductions. It would be of interest to investigate the behavior of regularized  solutions numerically.

\section*{Acknowledgements}

We thank B. Dubrovin, K Khusnutdinova,  B. Konopelchenko, A. Mikhailov, V. Novikov,  M. Pavlov and I. Strachan for clarifying discussions.
This research   was  supported by the EPSRC grant EP/D036178/1,  the
European Union through the FP6 Marie Curie RTN project ENIGMA (Contract
number MRTN-CT-2004-5652), and the ESF programme MISGAM.

\end{document}